\documentclass[manuscript=article]{achemso}            		

\setkeys{acs}{usetitle=true}

\usepackage{graphicx}
\usepackage{subcaption}
\usepackage{multirow}
\usepackage[version=3]{mhchem}
\usepackage{longtable}
\usepackage{siunitx}
\usepackage{hyperref}
\usepackage{breakurl}
\usepackage{amsmath}
\usepackage{pdfpages}
\DeclareSIUnit\angstrom{\text{\normalfont\AA}}
\usepackage{xr}

\newcommand{\erf}[1]{\mathrm{erf}\left(#1\right)}

\title{A Linear-Scaling, Charge-Aware Foundation Potential for Atomistic Simulations}
\author{Tsz Wai Ko}
\email{t1ko@ucsd.edu}
\affiliation[UCSD]{Aiiso Yufeng Li Family Department of Chemical and Nano Engineering, University of California San Diego, 9500 Gilman Dr, Mail Code 0448, La Jolla, CA 92093-0448, United States}
\author{Runze Liu}
\affiliation[UCSD]{Aiiso Yufeng Li Family Department of Chemical and Nano Engineering, University of California San Diego, 9500 Gilman Dr, Mail Code 0448, La Jolla, CA 92093-0448, United States}
\author{Adesh Rohan Mishra}
\affiliation[UCSD]{Aiiso Yufeng Li Family Department of Chemical and Nano Engineering, University of California San Diego, 9500 Gilman Dr, Mail Code 0448, La Jolla, CA 92093-0448, United States}
\author{Zihan Yu}
\affiliation[UCSD]{Aiiso Yufeng Li Family Department of Chemical and Nano Engineering, University of California San Diego, 9500 Gilman Dr, Mail Code 0448, La Jolla, CA 92093-0448, United States}
\author{Ji Qi}
\affiliation[UCSD]{Aiiso Yufeng Li Family Department of Chemical and Nano Engineering, University of California San Diego, 9500 Gilman Dr, Mail Code 0448, La Jolla, CA 92093-0448, United States}
\author{Shyue Ping Ong}
\email{ongsp@ucsd.edu} 
\affiliation[UCSD]{Aiiso Yufeng Li Family Department of Chemical and Nano Engineering, University of California San Diego, 9500 Gilman Dr, Mail Code 0448, La Jolla, CA 92093-0448, United States}
\date{}

\usepackage{lineno}
\begin{document}
\maketitle
\nolinenumbers

\begin{abstract}
Electrostatics govern charge transfer and reactivity in materials. However, most foundation potentials (FPs) either neglect explicit electrostatic interactions or come at prohibitive computational cost. Here, we introduce charge-equilibrated TensorNet (QET), an equivariant, charge-aware architecture that achieves linear scaling with system size via an analytically solvable charge-equilibration scheme. We demonstrate that a trained QET FP matches state-of-the-art FPs on materials property benchmarks but delivers qualitatively different predictions in systems dominated by electrostatic interactions. The QET FP reproduces the correct structure and density of the NaCl–\ce{CaCl2} ionic liquid and the crystallization of  \ce{Ge1Sb2Te4} phase change memory, which charge-agnostic FPs miss. We further show that a fine-tuned QET captures reactive processes at the Li/\ce{Li6PS5Cl} solid-electrolyte interface and supports simulations under applied electrochemical potentials. These results remove a fundamental constraint in large-scale atomistic simulations of electrostatics and establish a general, data-driven framework for charge-aware FPs with transformative applications in energy storage, catalysis, and beyond.
\end{abstract}
		
\section{Introduction}

Modeling electrostatics is essential to describe charge transfer, reactivity, and non-bonded interactions in molecules and materials. Although ab initio methods such as density functional theory (DFT) can accurately capture these interactions, their prohibitive computational cost and poor scaling (typically $\geq \mathcal{O}(N_e^3)$, where $N_e$ is the number of electrons) render them impractical for large length and long time scale simulations.  

Machine learning interatomic potentials (MLIPs)\cite{ko2023recent} represent a modern approach to overcome the accuracy-cost tradeoffs in traditional simulation approaches. MLIPs utilize an ML model to reproduce the potential energy surface (PES) from ab initio training data, providing accurate energy, force, and stress predictions at a fraction of the computational cost and linear scaling with respect to the number of atoms ($\mathcal{O}(N_{atoms})$, where $N_{atoms}$ is the number of atoms). A particularly exciting development in recent years is the concept of foundation potentials (FPs),\cite{chen2022universal} which are universal MLIPs that have comprehensive coverage of the periodic table. With rapid progress,\cite{deng2023chgnet,batatia2022mace,batzner2022,lysogorskiy2025graph} pre-trained FPs today can serve as surrogates for expensive DFT calculations in many applications, most notably in materials discovery where the ability to rapidly screen vast chemical spaces is a key requirement.

However, most MLIPs and FPs are limited in their ability to describe electrostatic interactions. Although graph-based MLIPs, such as the Materials 3-body Graph Network (M3GNet),\cite{chen2022universal} Message-passing Atomic Cluster Expansion (MACE),\cite{batatia2022mace} and Graph Atomic Cluster Expansion (GRACE)\cite{lysogorskiy2025graph}, can in principle extend the range of interactions by increasing the number of message-passing steps, this approach is inefficient and leads to a computationally prohibitive increase in the number of model parameters. Furthermore, increasing the number of interaction blocks may degrade the accuracy of the model due to over-smoothing and over-squashing.\cite{giraldo2023trade} Despite its acronym, it should be noted that the Crystal Hamiltonian Graph Network (CHGNet)\cite{deng2023chgnet} architecture does not capture electrostatics but rather regularizes model training using magnetic moments as a proxy for the oxidation state.

A common approach to address this long-standing challenge is to incorporate the charge equilibration (Qeq) scheme of \citet{rappe1991charge} into MLIPs\cite{ko2021general, kocer2022neural}. The general concept is to learn environment-dependent atomic electronegativities and hardnesses to distribute charges across the entire system so that the Qeq energy expression is minimized. These charges can then be used to calculate electrostatic interactions using Coulomb's law. Over the past few years, Qeq-based MLIPs, such as the fourth-generation high-dimensional neural network potential (4G-HDNNP)\cite{ko2021fourth, ko2023accurate}, atoms-in-molecules neural network with neural spin equilibration (AIMNet-NSE)\cite{anstine2025aimnet2}, and others\cite{ghasemi2015interatomic, jacobson2022transferable, maruf2025equivariant, fuchs2025learning}, have been developed to describe non-local charge transfer and multiple charge states for applications such as protonation/deprotonation of molecules and electrocatalysis. Some alternative approaches, such as global attention\cite{frank2024euclidean}, maximally localized Wannier function centers (MLWCs)\cite{yue2021short}, latent Ewald summation (LES)-based methods\cite{kim2024learning} and latent message passing\cite{wang2024neural}, can also accurately describe electrostatic interactions without explicit charge training.

Despite these promising advances, developing an FP for materials that can accurately capture non-local charge transfer across the periodic table remains an open challenge. This is primarily due to the lack of large-scale reference datasets with reliable partial charges and higher-order multipoles. Modern PES-quality datasets, such as the Materials Potential Energy Surface (MatPES)\cite{kaplan2025foundational} and OMat24\cite{barroso2024open} datasets, currently do not provide this information. Additionally, most Qeq implementations determine the atomic charges by solving an augmented system of linear equations with a ($N_{\mathrm{atoms}}+1)\times (N_{\mathrm{atoms}}+1$), where the additional row and column arise from the Lagrange multiplier used to enforce the total-charge constraint and ($N_{\mathrm{atoms}}$) is the number of atoms. Because the matrix is generally dense, storing it requires $\mathcal{O}(N_{\mathrm{atoms}}^2)$ memory. A direct solution scales $\mathcal{O}(N_{\mathrm{atoms}}^3)$ though iterative solvers can reduce the practical scaling to approximately quadratic\cite{kocer2025iterative}. Furthermore, conventional Ewald summation scales as $\mathcal{O}(N_{\mathrm{atoms}}^{3/2})$ with  optimal choice of the splitting parameter and the real- and reciprocal-space cutoffs. Grid-based Ewald methods, such as particle-particle particle-mesh\cite{hockney2021computer} and particle-mesh Ewald\cite{darden1993particle} can achieve quasi-linear $\mathcal{O}(N_{\mathrm{atoms}}\log N_{\mathrm{atoms}}$) scaling. When these grid-based methods are combined with an iterative Qeq solver to evaluate the Coulomb matrix–vector products, the asymptotic cost of Qeq can, in principle, achieve quasi-linear scaling\cite{gubler2024accelerating}, although realizing this scaling requires favorable convergence behavior, effective preconditioning, and sophisticated implementation of the fast electrostatic solver. These approaches have enabled fixed-charge classical force fields to scale to very large systems, including million-atom simulations on modern high-performance computing architectures\cite{gecht2020mdbenchmark,kutzner2025scaling}. Similar acceleration strategies have also been developed for variable-charge and polarizable reactive force fields for large-scale reactive MD simulations containing tens to hundreds of thousands of atoms depending on the force field, implementation, hardware architecture, and charge-equilibration strategy\cite{o2019performance,leven2021recent,johansson2025lammps}. However, these algorithmic advances primarily reduce the cost of solving the Qeq equations themselves and do not eliminate the additional overhead associated with differentiating through the charge-equilibration procedure in Qeq-based MLIPs. In particular, the electrostatic forces and stresses for Qeq-based MLIPs involve additional derivatives, such as those of geometry-dependent electronegativities, hardnesses, or latent charges with respect to atomic positions and lattice vectors. These terms introduce additional computational costs and memory consumption, particularly in GNN-based MLIPs where the required derivatives are obtained by the backpropagation through the neural-network computation graph. As a result, the Qeq scheme still remains a practical bottleneck for large-scale simulations with MLIPs.

In this work, we introduce a linear-scaling, charge-aware equivariant FP based on a charge-equilibrated TensorNet\cite{simeon2023tensornet} (QET, pronounced as ``ket'') architecture. QET attains linear scaling with system size by replacing conventional electrostatic solvers with an analytically solvable charge-equilibration scheme, while maintaining high accuracy in the PES and predicted atomic charges. We demonstrate that QET captures charge transfer and multiple charge states across diverse model systems. To train QET, we constructed a large charge-informed dataset, hereafter referred to as MatQ, comprising configurations sampled from structural deformations and high-temperature \textit{ab initio} molecular dynamics (AIMD) simulations and spanning 86 elements of the periodic table. We show that the QET FP compares favorably with the state-of-the-art FPs\cite{lysogorskiy2025graph, equiformer_v2} in near-equilibrium property benchmarks. Finally, we show that QET can be efficiently fine-tuned for complex reactive environments, such as the electrode/solid-electrolyte interface in next-generation all-solid-state lithium-ion batteries.

\section{Results}

\subsection{QET architecture}
The QET architecture is based on TensorNet\cite{simeon2023tensornet}, which has been shown to achieve significantly improved parameter and computational efficiency compared to other equivariant graph deep learning architectures by using Cartesian rank-2 tensor atomic embeddings. The Qeq scheme has been extensively detailed in previous work \cite{rappe1991charge, ghasemi2015interatomic, ko2021fourth, anstine2025aimnet2}, and an overview is provided in the Supplementary Information. Solving the Qeq system of linear equations using naive methods such as QR or LU decomposition results in $\mathcal{O}(N_{atoms}^3)$ scaling. Additionally, for periodic systems, setting up these equations involves computing the electrostatic energy via an Ewald summation, which scales from $\mathcal{O}(N_{atoms}^{1.5})$ in the best case, to $\mathcal{O}(N_{atoms}^{2})$ in the worst case, depending on the choice of splitting parameters. Despite advances in optimization techniques\cite{essmann1995smooth}, it remains prohibitively expensive to train a Qeq FP using large datasets or to perform large-scale simulations. 

\begin{figure}[htp]
    \centering
    \includegraphics[width=1.0\linewidth]{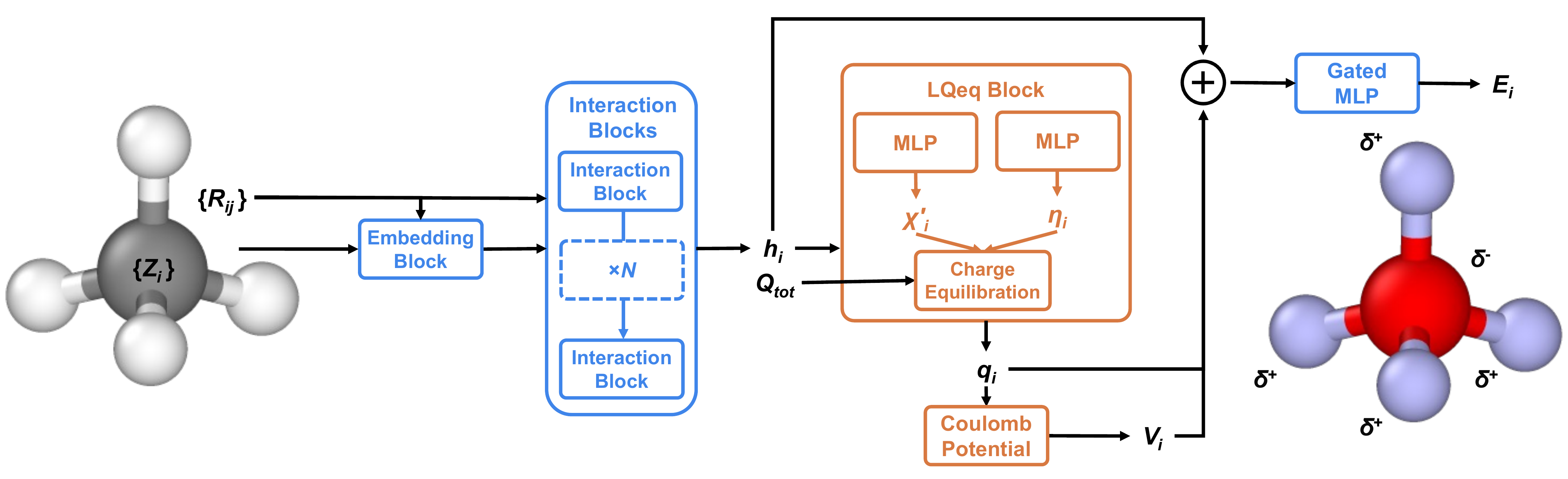}
    \caption{\textbf{QET architecture.} QET takes the atomic number $Z_i$ and atomic positions $R_i$ as inputs to an embedding block, which generates a set of rank-2 tensors representing atomic features. These features are subsequently refined through a series of interaction blocks to capture the latent representation of atomic environments. The graph-convoluted atomic features $h_i$ and total charge $Q_{\mathrm{tot}}$ are passed to the LQeq block to compute atomic charges. The LQeq block employs a two-step process to compute atomic charges. First, atomic features are fed into two distinct multi-layer perceptrons (MLPs) to predict environment-dependent electronegativities and hardnesses. To avoid undefined solutions in Eqn. \ref{eq:qeq_solution}, the softplus activation is used to ensure that the predicted hardness values are positive. These quantities are then used to calculate globally distributed charges, constrained by the total charge, using an analytical charge equilibration scheme. The charges are then used to calculate the Coulomb potential $V_{i}$ acting on the central atom arising from neighboring atoms. Finally, the atomic features, charges, and Coulomb potential are concatenated and fed into a gated MLP to predict atomic energies $E_i = \phi(h_i\oplus q_{i}\oplus V_{i})$, which are in turn summed to obtain the total energy.\label{fig:QET_architecture}}
    
\end{figure}

The schematic of the QET architecture is shown in Fig.~\ref{fig:QET_architecture}. To develop a linear-scaling charge equilibration (LQeq) scheme, we take inspiration from traditional mean-field approaches such as the Hartree-Fock method, but modernized to exploit machine learning approaches to mapping non-linear relationships. We introduce an effective electronegativity $\chi_i'$\cite{jacobson2022transferable,wang2024espalomacharge}, defined as follows:
\begin{equation}\label{eqn:qet1}
    \chi_i' = \chi_i + \frac{q_{i}}{2 \sigma_i \sqrt{\pi}} + \sum_{i>j}^{N_{\mathrm{atoms}}} \frac{\erf{\frac{R_{ij}}{\sqrt{2(\sigma_i^{2} + \sigma_j^{2})} }}}{R_{ij}} q_{j},
\end{equation}
Here, $\chi_i$ is the electronegativity in the standard Qeq scheme, $\sigma_i$ and $\sigma_j$ denote the width of the Gaussian-distributed charges, $q_i$ represents the partial charge of atom $i$, and $R_{ij}$ is the distance between atoms $i$ and $j$.

The partial charges $q_i$ are then given by an elegant analytic solution in terms of the effective electronegativities $\chi_i'$ and hardnesses $\eta_i$, as follows:
\begin{equation}\label{eq:qeq_solution}
    q_i = -\eta_i^{-1} \left[\chi_i' -\frac{Q_{\mathrm{tot}}+\sum_{i=1}\chi_i'\eta_i^{-1}}{\sum_{i=1}\eta_{i}^{-1}} \right]
\end{equation}
A complete derivation of this analytical solution is provided in the Supplementary Information. 

The main difference between LQeq and standard Qeq is in the treatment of interactions between fragments beyond the message passing cutoff. Because the effective electronegativities and hardnesses in LQeq are predicted from local atomic environments, they no longer respond to another fragment once the fragments are separated beyond the message-passing cutoff. As a result, LQeq does not introduce an artificial distance-dependent charge transfer between well-separated fragments, unlike standard QEq schemes where non-zero off-diagonal Coulomb couplings can lead to spurious long-range charge delocalization. However, the same feature also means that LQeq alone cannot explicitly describe genuine distance-dependent charge transfer between fragments beyond the cutoff, except through the global charge constraint. This limitation is partly mitigated in many condensed-phase systems by the nearsightedness of electronic matter, which makes direct charge transfer over such distances weak, although long-range electrostatic polarization may still remain important.

The Coulomb potential $V_i$ arising from neighboring atoms $N_{\mathrm{neig},i}$ is then calculated as:
\begin{equation}
         V_{i} = \sum_{j=1}^{N_{\mathrm{neig},i}}\frac{\mathrm{erf}(\frac{R_{ij}}{\sqrt{2(\sigma_i^{2} + \sigma_j^{2})}})q_{j}}{R_{ij}}f_{\mathrm{cut}}(R_{ij}) \quad 
\end{equation}
where $f_\mathrm{cut}$ is a cutoff function that ensures that the Coulomb potential contributions and its derivatives smoothly decay to zero at distance $R_c$. Finally, the atomic features, charges, and Coulomb potential are concatenated and fed into a gated multi-layer perceptron (MLP) to predict atomic energies. 

As shown in Fig. S1, QET incurs only a marginal increase in inference cost and exhibits scaling comparable to the standard non-charge-informed TensorNet (hereafter ``TensorNet''). In contrast, TensorNet combined with the conventional Qeq scheme exhibits prohibitive non-linear scaling, rendering it impractical for simulations of realistic system sizes.

It should be noted that the default QET framework does not incorporate explicit long-range electrostatic interactions to keep strictly linear computational scaling with respect to the number of atoms. QET can be still complemented by Ewald summation using either supervised or latent charges if necessary.

\subsection{Architecture benchmarks}

We benchmark QET using four datasets that have been widely used to evaluate the performance of MLIPs in modeling different charge states and charge transfer in different applications:\cite{ko2021fourth} (1) a long carbon chain terminated with hydrogen atoms (C$_{10}$H$_2$/C$_{10}$H$_3^+$); (2) a silver trimer with total charge of $\pm 1$ (\ce{Ag3^{+/-}}); (3) an ionized Na$_9$Cl$_8^+$ cluster with a removed Na atom; and (4) gold dimer adsorption on MgO(001) surface with and without Al doping. MLIPs that do not incorporate atomic charges  struggle to achieve reliable energy and force errors in these systems. 
The trained QET models compare favorably to three charge-informed MLIPs trained on the same datasets (see Table S1). The QET models achieve root mean squared errors (RMSEs) in energies less than 1 meV atom$^{-1}$ for all datasets except Ag$_3^{+/-}$ and RMSEs in forces less than 20 meV \si{\angstrom}$^{-1}$ across all datasets. It should be noted that both 4G-HDNNP and CELLI suffer from the $\mathcal{O}(N_{atoms}^3)$ scaling of the traditional Qeq scheme, while CACE-LR still requires a best-case $\mathcal{O}(N_{atoms}^{1.5})$ calculation of electrostatic interactions using the Ewald method. In contrast, QET scales as $\mathcal{O}(N_{atoms})$. A crucial observation is that the charges predicted by the LQeq scheme are comparable to those from the conventional Qeq scheme in 4G-HDNNP and CELLI as well as CACE-LR based on latent environment dependent charges, demonstrating that the LQeq approximation (Eqn.~\ref{eqn:qet1}) does not result in a loss of accuracy in predicted charges and PES properties. 

\subsection{Importance of electrostatics}

Here, we illustrate that incorporating electrostatics leads to not only quantitative but qualitative differences in the simulated behavior of the model systems.

\begin{figure}
    \centering
    \includegraphics[width=1.0\linewidth]{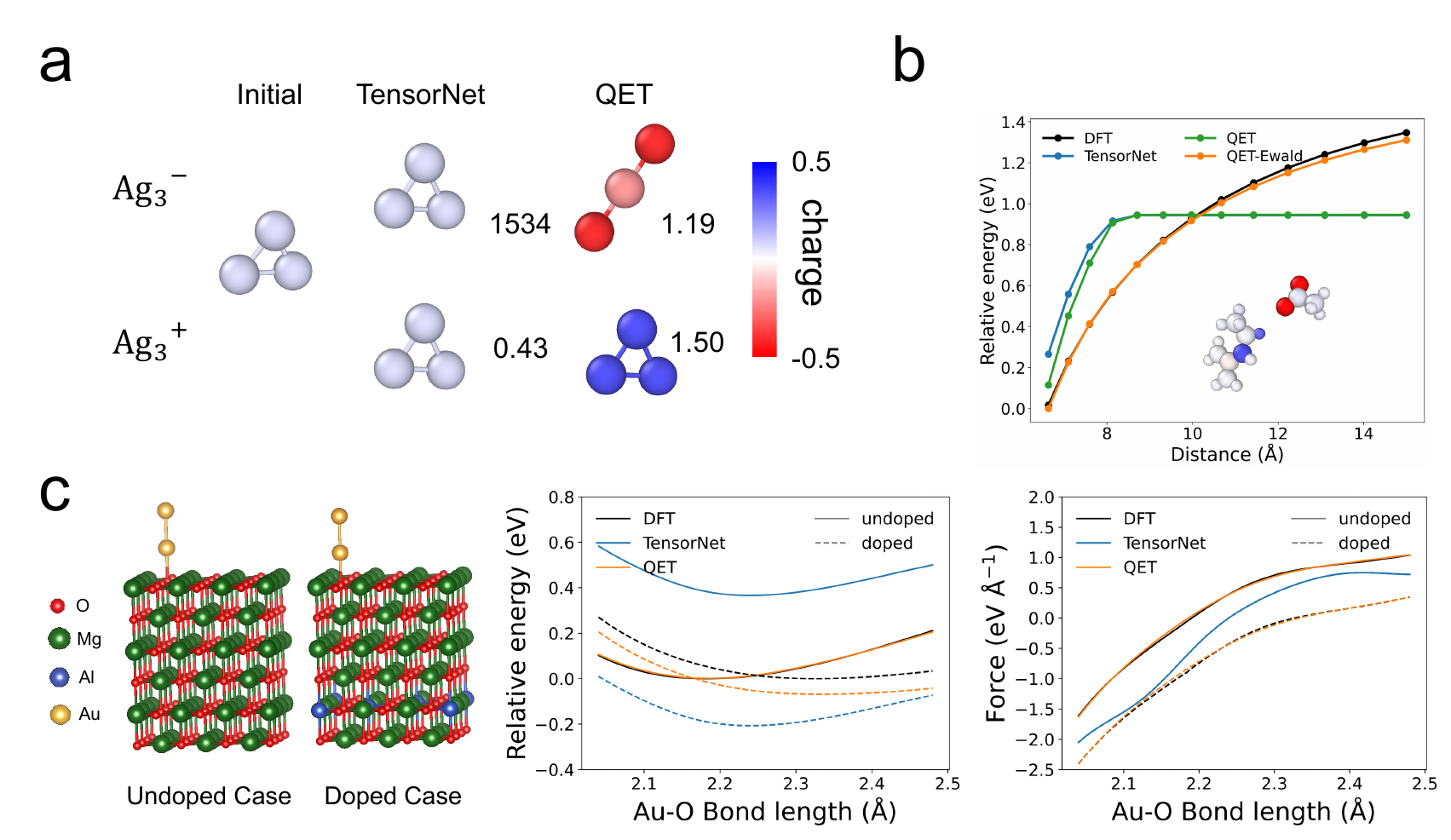}
    \caption{\textbf{Ablation studies of QET on charged systems and long-range electrostatic interactions.} \textbf{a}, Modeling charge-dependent equilibrium geometries in \ce{Ag3+}/\ce{Ag3-}. \textbf{b}, Binding-energy curves of the \ce{C3N3H10+}/\ce{C2O2H3-} molecular dimer. \textbf{c}, Relative energy and force profiles of an Au dimer adsorbed on undoped and Al-doped MgO(001). 
The numbers in panel \textbf{a} indicate the root mean square deviation (RMSD) in 10$^{-3}$\si{\angstrom} with respect to the DFT-relaxed geometries. The color coding of the Ag clusters and the molecular dimer represents the partial charges predicted by QET ranging from -0.5 to 0.5 e. The energies shown in panel \textbf{b} and \textbf{c} are relative to the lowest DFT value. }
    \label{fig:ablation_studies}
\end{figure}

Fig. \ref{fig:ablation_studies}a shows the relaxed geometries of \ce{Ag3^{+/-}} using QET and TensorNet. All geometry relaxations were initialized from the same perturbed DFT-relaxed \ce{Ag3^{+}} structure. The QET-relaxed structures for \ce{Ag3^+} and \ce{Ag3^-} have very small root mean square deviations (RMSD) of $1.50 \times 10^{-3}$ $\si{\angstrom}$ and $1.19 \times 10^{-3}$ $\si{\angstrom}$, respectively, relative to the DFT-relaxed structures. In contrast, TensorNet predicts identical equilibrium geometries for both charge states because it does not explicitly incorporate global charge information. As a result, it fails to reproduce the DFT-relaxed geometry of \ce{Ag3^-}, yielding a much larger RMSD of 1.534 \si{\angstrom}. 
QET predicts that the energy of Ag$_3^{-}$ is lower than that of Ag$_3^{+}$ by 7.867 eV, which is in excellent agreement with the DFT value of -7.865 eV. In contrast, TensorNet shows no energy difference between two charge states. The QET-predicted charges are also within 1 m$e$ of the DFT-Hirshfeld charges in both cases. This highlights the role of the LQeq scheme in QET, which determine the partial charges according to the given total charge. The resulting charges provide distinct inputs to the atomic-energy readout, enabling QET to distinguish different charge states within a single model.

Fig. \ref{fig:ablation_studies}b shows the binding-energy curves of the charged \ce{C3N3H10+}/\ce{C2O2H3-} dimer, which was used as a benchmark in the CACE-LR study\cite{king2025machine}. TensorNet fails to reproduce the intermolecular interactions beyond the message-passing cutoff ($R_\mathrm{c}=5$ \si{\angstrom}), as no messages are exchanged between the two well-separated fragments. QET slightly improves the interaction energies at shorter separations through its charge-aware representation, but likewise fails to capture the long-range interaction beyond the cutoff. In contrast, QET combined with Ewald summation using latent charges accurately reproduces the DFT binding-energy curve over the entire distance range up to 15 \si{\angstrom}, demonstrating that QET can be naturally extended with explicit long-range electrostatics when required. The corresponding latent charges are also physically meaningful, with total charges of approximately +0.82e and -0.82e for \ce{C3N3H10+} and \ce{C2O2H3-}, respectively. Moreover, the LQeq scheme exactly enforces the overall charge neutrality of the dimer without any artificial post-processing, such as subtracting the mean atomic charge, which is commonly adopted in purely environment-dependent latent-charge models.

Fig.~\ref{fig:ablation_studies}c compares the QET, TensorNet and DFT calculated energies as a function of the Au-O bond length in the undoped and doped \ce{Au2}-MgO(001) system. DFT calculations predict two distinct energy and force curves for the undoped and doped cases with respect to the distance between the Au dimer and the nearest oxygen atom. The equilibrium bond length predicted by DFT for the undoped case is approximately 2.19 \si{\angstrom}, while that for the Al-doped case is approximately 2.33 \si{\angstrom}. The QET energy and forces curves are in excellent agreement with the reference DFT curves, and the equilibrium bond lengths are within 0.01 \si{\angstrom} of the DFT values. In contrast, TensorNet predicts the same energy curves with a constant offset and the same force acting on the Au dimer with an equilibrium distance of approximately 2.24 \si{\angstrom} for both doped and undoped cases. Additional evidence of the qualitative difference between QET and TensorNet can be seen in the principal component analysis (PCA) of the atomic features (Fig. S2). The TensorNet atomic features fail to distinguish between the undoped and Al-doped cases, while the QET atomic features exhibit two distinct clusters for one of the Au atoms for the undoped and doped surfaces. The performance of QET augmented with Ewald summation using either supervised or latent charges on the test sets of these benchmark datasets are summarized in Table S2. Overall, these studies demonstrate that the proposed QET architecture effectively captures non-local charge transfer and accurately describes charge-dependent interactions, achieving performance comparable to QET with explicit Ewald summation in condensed systems. For systems containing well-separated fragments beyond the effective receptive field, QET can be naturally complemented by Ewald summation to recover explicit long-range electrostatic interactions.

\subsection{Charge-informed foundation potential} 

To train a charge-informed QET FP, we generated a Materials Charge (MatQ) dataset, which comprises around 120,000 structures obtained by deforming Materials Project crystals\cite{horton2025accelerated}, and 1000K/3000K high-temperature AIMD trajectories in the OMat24\cite{barroso2024open} validation set. This MatQ data set covers 86 elements and is computed using well-converged DFT static calculations with the Perdew-Burke-Ernzerhof (PBE) functional\cite{perdew1996generalized}. The atomic charges are computed with the density derived electrostatic and chemical (DDEC6) electron density partitioning\cite{manz2016introducing} (see Methods section). The distribution of elements, PES properties and charges of the dataset are presented in Fig. S3 and S4.

Fig. S5 and S6 show the parity plots of the PES and charge errors for the QET and TensorNet FPs trained on the MatQ dataset. The test MAEs in energies, forces and stresses of the QET-MatQ FP are 30.6 meV atoms$^{-1}$, 103 meV \si{\angstrom}$^{-1}$ and 0.439 GPa, while those of the TensorNet-MatQ FP are 29.8 meV atoms$^{-1}$, 107 meV \si{\angstrom}$^{-1}$ and 0.463 GPa, respectively. These MAEs are in line with those of previous state-of-the-art FPs. The MAE of the charges of the QET-MatQ FP is approximately 26.8 m$e$, which is only slightly larger than the MAEs for the custom QET MLIPs for the model systems in Table S2. 

\subsubsection{Equilibrium and non-equilibrium properties}

We benchmarked the QET and TensorNet FP trained using the MatQ dataset (QET-MatQ and TensorNet-MatQ) against three state-of-the-art FPs - TensorNet trained on the MatPES dataset (TensorNet-MatPES-PBE-v2025.2),\cite{kaplan2025foundational} Graph Atomic Cluster Expansion trained on OMat24 and then fine-tuned on the subset of OMat24, Alexandria\cite{schmidt2024improving} and MPTraj\cite{deng2023chgnet} (GRACE-2L-OAM)\cite{bochkarev2024graph} and the non-conservative EquiformerV2 trained on OMat24 (EqV2-L-OMat24)\cite{liao2024equiformerv}. We explicitly denote each FP using a ``architecture-dataset-version'' format. It has been shown that the training dataset has an overwhelming effect on the accuracy of the resulting FP, while the specific architecture plays a relatively minor role.\cite{kaplan2025foundational} All three benchmarked FPs are trained on much larger datasets - MatPES ($\sim 400,000$ structures) and OMat24/OAM ($\sim 100$ million structures) - compared to the QET FP (MatQ with $\sim 120,000$ structures). 

From Table.~\ref{tab:foundation-potentials}, we may observe that the performance of the QET-MatQ FP is in line with our expectations based the training data set. There is a negligible difference in performance between the QET and TensorNet FPs trained on the same MatQ dataset, as well as the TensorNet and QET trained on the MatPES dataset. The slightly better performance of the GRACE-2L-OAM FP is likely attributed to the larger amount of training data for near-equilibrium and non-equilibrium structures and the higher model complexity. The non-conservative EqV2-L-OMat24 only performs well in structure relaxations ($d$) and heat capacities but significantly worse in mechanical properties and thermal conductivities due to the discontinuity in PES caused by direct force prediction. 

Additionally, we performed studies to assess the effect of explicitly including long-range electrostatic interactions on Li-ion migration barriers and ionic conductivities in battery materials. QET-Ewald retains the same QET architecture but includes an additional Ewald-summation energy term evaluated using the QET-predicted charges. Starting from the pretrained QET FP, we fine-tuned QET-Ewald FP on the same MatQ dataset. This comparison therefore isolates the contribution of explicit long-range electrostatics. Fig. S7 shows that QET and QET-Ewald achieve comparable accuracy for the migration barriers in the nebDFT2K benchmark~\cite{dembitskiy2025benchmarking}. Including the Ewald-summation term reduces the MAE only marginally, from 0.19 to 0.18~eV. Similarly, on the MVL-batt ionic conductivity benchmark for 172 battery materials\cite{kaplan2025foundational} as shown in Fig. S8, the MAE in the logarithmic ionic conductivity, defined as $\log_{10}[\sigma/(\mathrm{mS\,cm^{-1}})]$, decreases only slightly from 0.23 for QET to 0.22 for QET-Ewald. These results suggest that, for most solid electrolytes represented in these benchmarks, QET already captures the major electrostatic interactions governing Li-ion migration and transport through the truncated Coulomb potentials arising from neighboring charges. Consequently, including an additional long-range Ewald-summation term provides only a modest improvement.

\begin{table}[htp]
\centering
\caption{\textbf{Equilibrium and non-equilibrium property benchmarks for foundation potentials.} Reported values correspond to the average fingerprint distance between the FP-relaxed and DFT-relaxed crystals $d$, the mean absolute errors (MAEs) of the bulk modulus $K_{\mathrm{vrh}}$ (GPa), shear modulus $G_{\mathrm{vrh}}$ (GPa), heat capacity $C_{\mathrm{v}}$ (J mol$^{-1}$K$^{-1}$), thermal conductivity $\kappa$ (log$_{10}$ Wm$^{-1}$K$^{-1}$) and Li-ion migration barriers $E_\mathrm{m}$ (eV). The test data for geometry relaxations, elastic, phonon properties and migration paths were obtained from Ref~\citenum{wangpredicting2021}, ~\citenum{dunn2020benchmarking}, ~\citenum{loew2025universal} and ~\citenum{dembitskiy2025benchmarking}, respectively, as detailed in Ref.~\citenum{kaplan2025foundational} and Method section. The numbers in parentheses denote the number of training parameters (in millions).}
\label{tab:foundation-potentials}
\begin{tabular}{c|cccccc}
\hline
FP & $d$ & $K_{\mathrm{vrh}}$  &  $G_{\mathrm{vrh}}$  & $C_{\mathrm{v}}$  & $\kappa$  & $E_\mathrm{m}$   \\
\hline
EqV2-L-OMat24 (153M) & 0.17 & 31 & 22 & 9 & 0.95 & 0.3 \\
GRACE-2L-OAM (26.4M) & 0.23  & 14 &  17 & 7 & 0.06 & 0.17 \\
TensorNet-MatPES (0.84M) & 0.36 & 19 & 16  & 27 & 0.18 & 0.18 \\
QET-MatPES (0.905M) & 0.37 & 17 & 15 & 16 & 0.16 & 0.15 \\
TensorNet-MatQ (0.27M) & 0.38 & 17 & 17  & 15 & 0.17 & 0.17 \\
QET-MatQ (0.29M) & 0.36 & 19 & 19 & 16 & 0.20 & 0.19 \\
\hline
\end{tabular}
\end{table}

\subsubsection{NaCl-\ce{CaCl2} ionic liquid}

To illustrate the value of a charge-equilibrated FP, we performed simulations of the well-studied NaCl-\ce{CaCl2} ionic liquid system with 1040 atoms using QET-MatQ, TensorNet-MatQ and GRACE-2L-OAM. We find that the QET-MatQ FP slightly outperforms the TensorNet-MatQ FP in the predicted energies, forces, and stresses of a  training set containing about 800 NaCl-\ce{CaCl2} ionic liquid structures taken from Ref.\citenum{guo2023al4gap} recomputed using consistent DFT settings. The MAEs in energies for QET-MatQ and TensorNet-MatQ are 12 meV atom$^{-1}$ and 17 meV atom$^{-1}$, respectively, while the MAEs in forces are 65 and 66 meV \si{\angstrom}$^{-1}$, respectively. Furthermore, the validation charge MAE of QET is 0.017 e with respect to DFT DDEC6 charges. These results demonstrate that QET FP preserves the high accuracy of the underlying TensorNet architecture while providing an explicit and physically meaningful description of the charge distribution.

\begin{figure}[htp]
    \centering
    \includegraphics[width=1.0\linewidth]{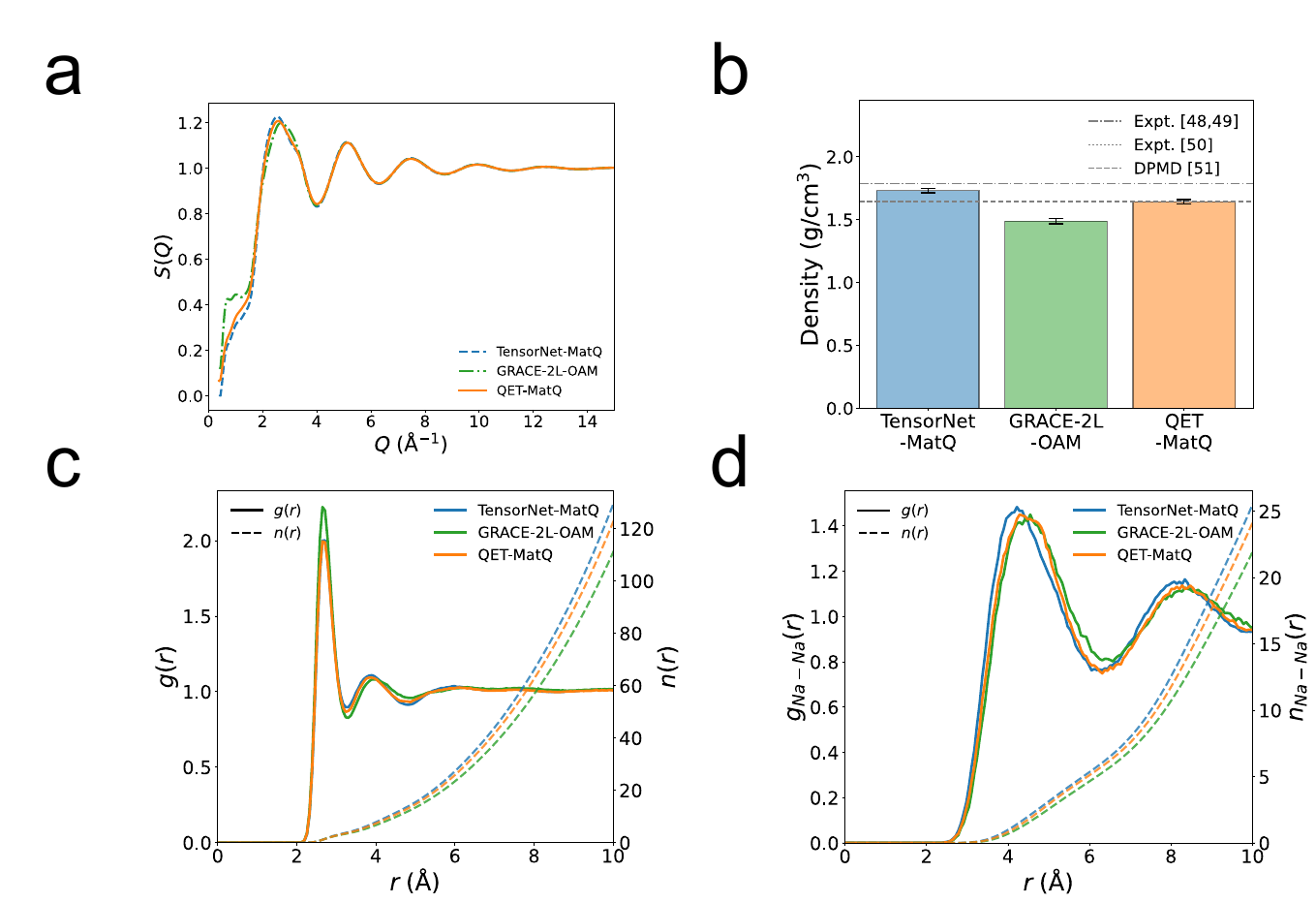}
    \caption{\textbf{Simulated properties of NaCl-\ce{CaCl2} ionic liquid.} \textbf{a,} Calculated structure factor and \textbf{b,} averaged densities of NaCl-\ce{CaCl2} obtained from 100 ps \textit{NPT} MD simulations at 1200 K using the QET-MatQ, TensorNet-MatQ and GRACE-2L-OAM FPs. The experimental density was obtained from Ref.~\cite{janz1988thermodynamic,janz1979physical,tian2021thermal} and DPMD predicted density was obtained from Ref.~\cite{gegentana2024deep}. \textbf{c}, Total and \textbf{d}, Na-Na radial distribution functions together with the corresponding cumulative coordination.}
    \label{fig:ionic_liquid}
\end{figure}

MD simulations using the QET-MatQ, TensorNet-MatQ, and GRACE-2L-OAM FPs predict distinct structural properties for the NaCl--\ce{CaCl2} ionic liquid. As shown in Fig.~\ref{fig:ionic_liquid}a, the calculated structure factors from all three models are in excellent agreement for $Q > 2$~\si{\angstrom^{-1}}, with only a slight shift in the first diffraction peak near $Q\approx2.5$~\si{\angstrom^{-1}}. This indicates that all three models describe the short-range liquid structure similarly. In contrast, noticeable differences emerge in the low-$Q$ region ($Q < 2$~\si{\angstrom^{-1}}), where GRACE-2L-OAM exhibits an enhanced scattering intensity relative to TensorNet-MatQ and QET-MatQ. We interpret this low-$Q$ enhancement as arising from differences in long-range packing and density fluctuations. This interpretation is supported by the predicted densities and coordination statistics. Fig.~\ref{fig:ionic_liquid}b shows that the average density predicted by QET-MatQ is in excellent agreement with both experiment~\cite{janz1988thermodynamic,janz1979physical} and the customized MLIP\cite{gegentana2024deep}, whereas TensorNet-MatQ and GRACE-2L-OAM overestimate and underestimate the liquid density, respectively. Consistent with this trend, the total radial distribution functions (RDFs) in Fig.~\ref{fig:ionic_liquid}c are nearly identical within the first few coordination shells, indicating similar short-range atomic arrangements. However, the cumulative coordination number of GRACE-2L-OAM becomes systematically lower at larger distances, reflecting a more open atomic packing and lower average density. The Na-Na partial RDF and cumulative coordination number (Fig.~\ref{fig:ionic_liquid}d) exhibit the same trend, suggesting that the primary differences between the models originate from intermediate- and long-range packing rather than the local coordination environment. The remaining partial RDFs and cumulative coordination numbers are provided in Fig. S9.

\subsubsection{\ce{Ge1Sb2Te4} phase change memory}
\begin{figure}[htp]
    \centering
    \includegraphics[width=1.0\linewidth]{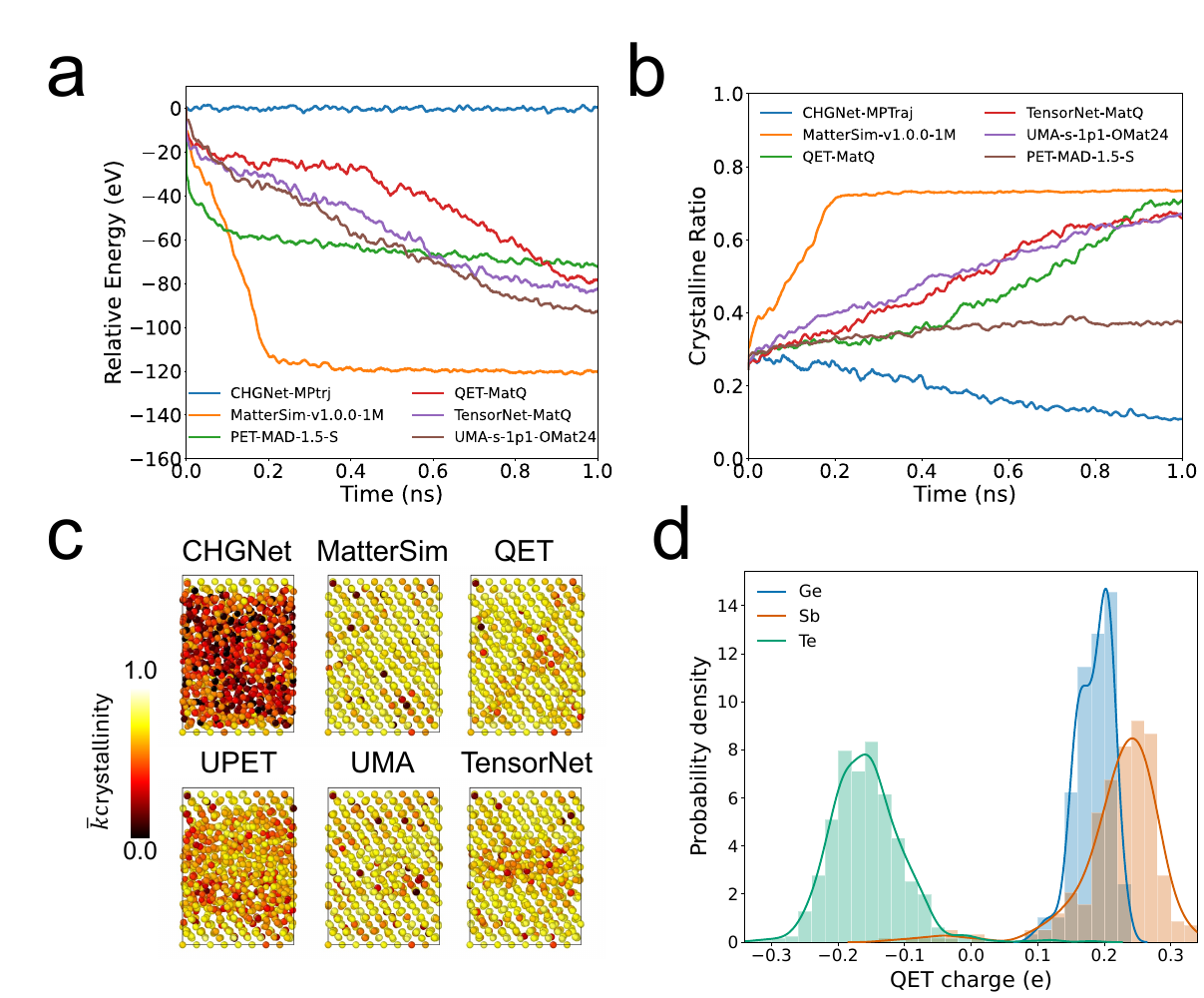}
    \caption{\textbf{Simulated properties of \ce{Ge1Sb2Te4} phase-change memory.}
    \textbf{a,} Time evolution of the relative potential energy and
    \textbf{b,} crystalline ratio obtained from 1 ns \textit{NVT} MD simulations at 600 K using different FPs.
    \textbf{c,} Final snapshots colored by the SOAP-based crystallinity metric $\bar{k}$.
    \textbf{d,} Distribution of QET-predicted atomic charges for Ge, Sb, and Te atoms for the final MD snapshot. The potential energy of each FP is reported relative to its value at $t=0$ ns.}
    \label{fig:gst}
\end{figure}

We conducted an additional ablation study for the QET-Mat FP to demonstrate how global charge distribution improves the description of crystallization behavior of phase-change materials. we performed 1 ns \textit{NVT} simulations at 600 K for amorphous \ce{Ge1Sb2Te4} using QET-MatQ together with several recent FPs\cite{deng2023chgnet,yang2024mattersim,wood2026family,malosso2026high} trained on different databases. Following Ref.~\citenum{xu2022unraveling}, we quantified crystallization using a per-atom crystallinity metric, $\bar{k}$, which measures the similarity between each atom's local environment and an ensemble of reference crystalline environments using the smooth overlap of atomic positions (SOAP) descriptor\cite{bartok2013representing}.

Fig~\ref{fig:gst}a and b show the evolution of the relative potential energy and crystalline ratio during crystallization. MatterSim exhibits the fastest crystallization, reaching a crystalline ratio of approximately 0.73 at about 200 ps, accompanied by the largest decrease in relative potential energy of around 120 eV. In contrast, CHGNet becomes more disordered throughout the simulation, with little change in relative potential energy. QET-MatQ, TensorNet-MatQ, UMA-s-1p1-OMat24, and PET-MAD-1.5-S predict distinct crystallization kinetics and potential energy profiles. QET-MatQ predicts a higher crystalline ratio than both TensorNet-MatQ and UMA-s-1p1-OMat24, reaching approximately 0.70 after 1 ns compared with approximately 0.67 for both models. The continuous decrease in relative potential energy predicted by QET-MatQ is consistent with the progressive crystal growth observed during the simulation. Finally, UPET-MAD-1.5-S shows only limited crystallization, with the crystalline ratio remaining below approximately 0.40 and only a modest decrease in relative potential energy.

Representative final configurations are shown in Fig.~\ref{fig:gst}c. Compared with TensorNet-MatQ and UMA-s-1p1-OMat24, QET-MatQ produces more extended crystalline domains with fewer residual amorphous regions, consistent with the larger crystalline ratio. These results suggest that incorporating global charge distributions through the LQeq scheme improves the description of the atomic interactions governing crystallization, enabling QET-MatQ to more accurately capture the phase-transition dynamics of \ce{Ge1Sb2Te4}. Finally, the QET-predicted atomic charges (Fig.~\ref{fig:gst}d) for the last MD snapshot are chemically reasonable, with Ge and Sb carrying positive partial charges and Te carrying negative partial charges, consistent with their expected oxidation states. Overall, these results indicate that QET-MatQ reproduces crystallization behavior that is comparable to, or in better agreement with, previous AIMD studies\cite{zhou2023device}, than state-of-the-art FPs, despite being trained on substantially less data and using substantially fewer trainable parameters.

\subsection{Reactive battery interface simulations}

To further demonstrate the utility of QET, we performed simulations of the reactive Li/\ce{Li6PS5Cl} interface with fine-tuned QET and TensorNet FPs. The Li/\ce{Li6PS5Cl} interface is of major interest in all-solid-state lithium-ion batteries, with an abundance of DFT and experimental data on the reaction products.\cite{deng2017data, xu2022enabling, chen2024construction, an2025observing} Fine-tuning was carried out on only the PES properties (without charges) using the dataset previously used to fit a custom DeepMD MLIP for this system.\cite{an2025observing} The pre-trained QET and TensorNet FPs without fine-tuning do not provide stable simulations of this interface due to the lack of interfacial and surface structures in the training data. The RMSEs of the fine-tuned QET and TensorNet models are provided in Table S3, and are significantly lower than those of the customized DeepMD MLIP.\cite{an2025observing}

\begin{figure}[htp]
    \centering
    \includegraphics[width=1.0\linewidth]{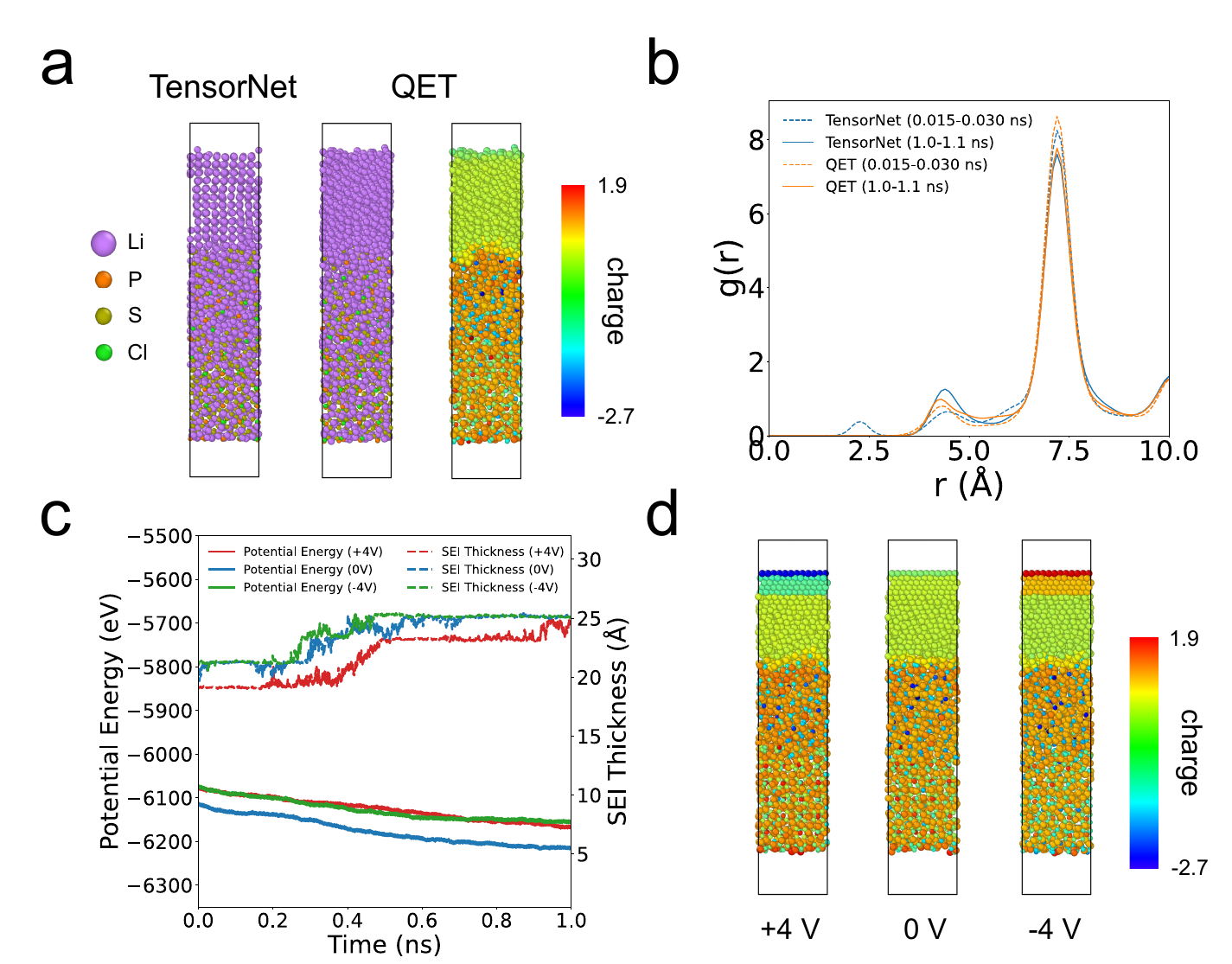}
    \caption{\textbf{Reactive simulations of the Li/\ce{Li6PS5Cl} interface.} \textbf{a,} Snapshots of the solid electrolyte interphase at t = 1.1 ns from \textit{NVT} MD simulations performed using fine-tuned QET and TensorNet potentials. \textbf{b,} Radial distribution function, calculated from the early-time interval from 0.015 to 0.03 ns of the trajectory. \textbf{c,} Potential energy and interphase thickness of the Li/\ce{Li6PSCl5} interphase obtained from the last 1 ns under applied potentials of $+4$ V, 0 V and $-4$ V. \textbf{d,} Comparison of the charge distribution of the Li/\ce{Li6P5SCl} at $t = 1.1$ ns for applied potentials of $+4$ V, 0 V and $-4$ V. All visualizations were performed using Ovito~\cite{stukowski2009visualization}. The heatmap colors indicate the partial charges predicted by QET.}
    \label{fig:comparison_SEI}
\end{figure}

Fig.~\ref{fig:comparison_SEI}a shows snapshots of the Li/\ce{Li6PS5Cl} interface after 1.1 ns of MD simulations at 300 K, consisting of a 0.1 ns equilibration period followed by a 1 ns production run using the fine-tuned QET-MatQ and TensorNet-MatQ FPs. A Langevin friction coefficient of $0.098~\mathrm{ps^{-1}}$ was used, corresponding to a damping timescale of approximately 10.2 ps. This relatively weak thermostat coupling provides effective temperature control while limiting artificial perturbations to Li-ion diffusion, interfacial reaction kinetics, and structural relaxation. In the QET simulations, the phosphorus atoms exhibit a wide range of charge values, consistent with their corresponding oxidation states as a \ce{P^{3-}} anion or as a \ce{PS4^{3-}} polyanion. The sulfur and chlorine atoms are consistently negatively charged, while the lithium atoms are either nearly neutral or positively charged. The separated charge distribution of Li, P, S, Cl atoms for the first and final snapshot of MD trajectory is shown in Fig. S10.  The P-P radial distribution function (RDF) is shown in Fig.~\ref{fig:comparison_SEI}b and those for Li-X (X = P, S, Cl) are provided in Fig. S11. Most of the peaks in the RDFs can be attributed to the formation of known interfacial reaction products such as \ce{Li3P}, \ce{Li2S}, \ce{LiCl}, etc. 
These analyses are used only to characterize local atomic correlations and identify bonding motifs associated
with interfacial reaction products, rather than to describe the full anisotropic structure of the interface.
In TensorNet simulations, a short P–P bond of 2.2 \AA~is observed during the equilibration ($t=0.015-0.03$ ns), which corresponds to the length of the P-P bond in elemental phosphorus. The formation of elemental phosphorus has never been observed in previous DFT or experimental studies\cite{chen2017ab}, and we speculate that this is the result of the lack of charge interactions in the TensorNet simulations. No such elemental P-P bonds were observed in the QET simulations. 
This eventually promotes the formation of phosphide-like structures within the solid electrolyte interphase (SEI), as reflected by the P–P RDF calculated over the final 0.1 ns of the production run. Notably, the P–P RDF is used only to assess local structural correlations and not to characterize the full anisotropic interfacial structure. 
A PCA analysis of the Li latent features in the QET simulations (Fig. S12) finds that distinct clusters corresponding to different local environments, e.g., in bcc Li metal, in bulk \ce{Li6PS5Cl} and experimentally known reaction products such as \ce{Li3P}, \ce{Li2S} and \ce{LiCl} at the interface.

We next demonstrate QET's ability to simulate interfacial reactions under an applied potential. Anodic and cathodic biases were imposed by shifting the electronegativities of Li atoms in the top three electrode layers to represent applied potentials of +4 V and -4 V, respectively, while the unbiased system (0 V) was included as a reference. The derivation of the constant-potential implementation within the LQeq framework is provided in the Supplementary Information. This approach is mathematically equivalent to adding an external electrostatic potential term to the charge-equilibration energy functional, thereby modifying the preferred charge distribution in response to an applied bias. The resulting charges are then obtained self-consistently subject to the global charge conservation constraint.

As shown in Fig.~\ref{fig:comparison_SEI}c, the applied potential modifies both the kinetics and extent of interphase growth. A cathodic bias (-4 V) accelerates interfacial reactions; the interphase thickens continuously and the potential energy decreases over the entire trajectory. Under a cathodic bias of (-4) V, the SEI thickness increases rapidly between approximately 0.25 and 0.5 ns before reaching a plateau near 25~\si{\angstrom}. The unbiased (0 V) simulation exhibits a more gradual, stepwise increase but reaches a similar final thickness at around 0.8 ns. In contrast, the (+4) V simulation shows delayed and less extensive SEI growth, with the thickness remaining lower over most of the trajectory and eventually reaching approximately 25~\si{\angstrom} after 1 ns.
The charge maps in Fig.~\ref{fig:comparison_SEI}d further illustrate the potential-dependent redistribution of charges across the electrode and interphase. As expected, the +4 V and -4 V simulations produce positively and negatively polarized electrodes, respectively, while the unbiased system lies between these two extremes. The QET-predicted charges on the Li, P, S, and Cl atoms under different bias potentials for the initial and final snapshots are shown in Fig. S13 and S14, respectively. Consistently, the RDF analysis in Fig. S15 shows that the short-range peaks associated with the reaction products \ce{Li3P}, \ce{Li2S}, and LiCl are preserved under all applied potentials, whereas the characteristic peaks of crystalline \ce{Li6PS5Cl} become slightly less pronounced under a -4 V bias, indicating enhanced decomposition of the solid electrolyte. Together, these results show that QET can accurately capture voltage-dependent interfacial chemistry in large-scale MD simulations.

\section{Discussion}

Incorporating electrostatic interactions is a crucial next step in the evolution of FPs. Here we demonstrate two key advances: a linear-scaling, charge-aware QET architecture and the training of a QET FP on a large, charge-informed dataset (MatQ).

A linear-scaling charge-equilibration scheme is essential for FPs to transcend conventional limits of length and time scales. This  effectively removes the long-standing electrostatic bottleneck in large-scale reactive simulations using charge-aware MLIPs.

Charge-aware FPs open a pathway to model phenomena governed by charge transfer. QET produces qualitatively distinct predictions across multiple benchmark systems and accurately captures the structure of the NaCl–\ce{CaCl2} ionic liquid and the crystallization of amorphous \ce{Ge1Sb2Te4}. The approach is particularly powerful for modeling reactive interfaces, such as electrode–electrolyte interfaces in batteries and adsorbate–catalyst interactions in heterogeneous catalysis, where charge redistribution drives functionality. Moreover, QET enables simulations under applied electrochemical potentials. 

QET can also be used to compute polarization and related electrical-response properties following the general strategy adopted in LES studies\cite{king2025machine,zhong2025machine}. Specifically, the environment-dependent charges predicted by QET can be used to evaluate the Berry-phase polarization within the modern theory of polarization\cite{resta1994macroscopic}. The Born effective charges (BECs) predicted by the QET-MatQ model using such an approach are in much better agreement with DFT reference values\cite{falletta2025unified} compared to those obtained with a TensorNet-LES model obtained by fine-tuning the pretrained TensorNet on MatQ after augmenting with Ewald-summation through latent charge (Fig. S16). We attribute this difference to the small latent charges learned by TensorNet-LES. The three message-passing layers already capture most interactions in the relatively small training structures, the explicit LES contribution is weakly constrained and yields only a small displacement-induced charge response. In contrast, the explicit charge supervision used in QET directly constrains the environment-dependent charges and their response to atomic displacements. Nevertheless, QET-MatQ still exhibits noticeable deviations from the DFT reference values, particularly for the strongly polar \ce{BaTiO3} system. Accurate and transferable predictions of BECs and related electrical-response properties will require explicit training against polarization, BEC, or dielectric-response data.

The present QET framework introduces a more flexible description of electrostatic interactions by using the LQeq-derived charges and the truncated Coulomb potential as descriptors for the atomic-energy readout, rather than representing the electrostatic energy solely through Coulomb interactions. Nevertheless, QET inherits some of the well-known limitations of QEq. In particular, delocalization and overpolarization errors may lead to unphysical charge redistribution and qualitatively incorrect responses under external electric fields, which are particularly relevant for heterogeneous interfaces. While the LQeq formulation partly mitigates these issues by eliminating the artificial long-range charge transfer associated with conventional Qeq, it remains an approximation and does not explicitly describe long-range polarization or higher-order electrostatic effects. Future developments could therefore combine the scalable QET framework with more advanced charge-equilibration schemes, such as QTP\cite{gergs2021generalized} or ACKS2\cite{verstraelen2013acks2}. In addition, QET could be further extended with learnable long-range kernels, such as those recently introduced in SOG-Net~\cite{ji2025machine,ji2026accurate}, to model a broader range of long-range interactions.

Furthermore, the introduction of external potential by shifting the electronegativities of fixed electrode atoms should be regarded as an efficient electrostatic approximation rather than a rigorous constant-potential electrochemical method. In contrast to grand-canonical or constant-potential approaches\cite{wang2025constant}, which explicitly control the electronic chemical potential through electron exchange with a reservoir, our model does not directly impose a Fermi level or allow the total number of electrons to vary. Instead, the applied bias enters as a perturbation to the atomic electronegativities, enabling voltage-induced charge redistribution and polarization while retaining the linear-scaling efficiency of the LQeq framework.

The recent MPNICE architecture\cite{weber2025efficient} and QET both aim to incorporate environment-dependent partial-charge redistribution into MLIPs using ML-predicted Qeq-like parameters. However, the two models differ in several central technical aspects. First, QET predicts both environment-dependent electronegativities and environment-dependent chemical hardnesses. Environment-dependent electronegativities determine whether atoms tend to become more positively or negatively charged in a given environment, while environment-dependent hardnesses control how easily the atomic partial charges can change during charge equilibration. By learning both quantities from the atomic environment, QET provides a more flexible Qeq parameterization for reproducing global partial-charge redistribution across diverse chemical environments. In contrast, MPNICE primarily introduces environment dependence through the predicted electronegativity, while the hardness is fixed using a radius-dependent expression based on atomic radii. Second, QET and MPNICE differ in how charge information is incorporated into the atomic-energy representation. MPNICE encodes bond distances and angles using modified atom-centered symmetry functions and feeds predicted partial charges back into the message-passing process through charge-weighted radial descriptor vectors. As a result, charge information directly modifies the latent atomic representations during the interaction updates. In contrast, QET is built on an equivariant TensorNet backbone, where graph-convoluted atomic features, Qeq-equilibrated partial charges, and the Coulomb potential arising from neighboring Gaussian charges are concatenated for atomic-energy prediction, without injecting charge information during message passing. This design preserves a cleaner functional separation between the equivariant geometric representation and the electrostatic charge-equilibration contribution, while still allowing global charge redistribution to influence the predicted atomic energies. We further developed an MPNICE-inspired variant, denoted QET-MPNICE, in which the LQeq-equilibrated partial charges are fed back into the TensorNet interaction blocks. Across three benchmark systems relevant to long-range electrostatics including Au dimer adsorbed on undoped and Al-doped MgO(001)\cite{ko2021fourth}, Copper on benzotriazole (Cu-BTA)\cite{maruf2025equivariant} and \ce{ LiCl-GaF3}\cite{king2025machine}, QET-MPNICE improves charge prediction while maintaining broadly comparable energy and force accuracy as shown in Table S4. However, introducing this charge-aware message-passing mechanism incurs a noticeable additional computational cost.

This work provides a proof of concept that an FP can be simultaneously accurate, charge-aware, and scalable. Further improvements in accuracy will come from expanding the diversity and fidelity of training data through broader community datasets such as MatPES\cite{kaplan2025foundational} or targeted fine-tuning for specific chemistries. Continued advances in architecture and algorithms will enhance parameter efficiency and inference speed. We urge the community to include atomic charges and magnetic moments in future open datasets, as such information is essential for developing more physically grounded FPs. With these collective efforts, charge-aware FPs like QET can unlock transformative advances across broad materials domains, including energy storage, catalysis, and beyond.

\section{Methods}

\subsection{Dataset construction}
The MatQ dataset comprises near-equilibrium and out-of-equilibrium structures. The 6,652,874 near-equilibrium structures comprise Materials Project crystals subject to strains of ±2\%, ±4\% and ±6\% along all crystallographic directions. The out-of-equilibrium structures were obtained from 1000K/3000K \textit{NVT/NPT} MD trajectories in the OMat24 validation set. The DImensionality-Reduced Encoded Clusters with sTratified (DIRECT) sampling approach\cite{qi2024robust} was then used to select 60,000 structures each from the near-equilibrium and out-of-equilibrium sets that comprehensively covers the configuration space with minimal overlap, significantly reducing the computational cost of reference data generation. 

Spin-polarized static DFT calculations were then performed using the Vienna Ab initio Simulation Package (VASP)\cite{kresse1996efficient}. The generalized gradient approximation (GGA) Perdew-Burke-Ernzerhof (PBE)\cite{perdew1996generalized} functional was employed to describe exchange-correlation interactions. The input files were generated using the ``MatPESStaticSet'' input set implemented in the Python Materials Genomics (pymatgen) package,\cite{ong2013pymatgen} which has been carefully benchmarked to ensure well-converged PES properties.\cite{kaplan2025foundational} The main parameters include a plane-wave kinetic energy cutoff of 680 eV, a $k$-spacing of 0.35 $\si{\angstrom}^{-1}$, and an electronic convergence condition of 10$^{-5}$ eV. DDEC6 charges were computed using Chargemol (version: 09\_26\_2017)\cite{limas2018introducing} from the DFT charge densities. 

The final MatQ dataset contains 114,445 structures, after excluding unconverged DFT calculations and structures with extremely large force components ($ |F_{x,y,z}| > 50$ eV \si{\angstrom}$^{-1}$). The fine-tuning dataset for Li/\ce{Li6PS5Cl} was obtained from Ref. \citenum{an2025observing}.

\section{Model Settings}
Unless otherwise specified, all models employed the following standard architecture. Interatomic distances were expanded using modified spherical Bessel basis functions\cite{kocer2019novel} constructed such that both their first and second derivatives vanish at the cutoff radius. Nine radial basis functions were used for this expansion. The SiLU activation function was used throughout the network to introduce nonlinearity. The widths of the Gaussian charge distributions were set to the covalent radii of the corresponding elements, and the cutoff radius used to define interatomic bonds was set to 5.0~\si{\angstrom}. The standard architecture comprised three interaction blocks, with 64 hidden channels in the embedding, interaction, and readout blocks.
For the ablation studies, a single interaction block was used for AuMgO and LiCl-GaF$_3$ systems. A smaller cutoff of 4~\si{\angstrom} was used in the ablation study of the Au–MgO system so that TensorNet could not capture the long-range charge transfer through message passing.
\subsection{Model training}

The datasets were split into 90\% for training and 10\% for validation. All model training was conducted using PyTorch Lightning. The AMSGrad variant of the AdamW optimizer was selected with a learning rate of $10^{-3}$ and a weight decay coefficient of $10^{-5}$. A cosine annealing scheduler adjusted the learning rate during training, with a maximum of $10^4$ iterations and a minimum learning rate of $10^{-5}$. The loss function for potential training is defined as:
\begin{equation}
    \mathcal{L}_{total} = w_{E}\mathcal{L}(E^{\mathrm{MLIP}}, E^{\mathrm{DFT}}) + w_{F}\mathcal{L}(F^{\mathrm{MLIP}},F^{\mathrm{DFT}}) + w_{\sigma}\mathcal{L}(\sigma^{\mathrm{MLIP}},\sigma^{\mathrm{DFT}}) + w_{q}\mathcal{L}(q^{\mathrm{MLIP}},q^{\mathrm{DFT}}),  
\end{equation}
where $E$, $F$, $\sigma$, and $q$ represent the total energy per atom, force components, stress tensor, and partial charge, respectively, and $\mathcal{L}$ is the $L$1 loss function with default PyTorch settings in this work.

For training the custom potentials for the four model systems, only charges, energies and forces were included in the loss function with weights $w_{E}=1.0$, $w_{F}=1.0$ and $w_{q}=0.1$ ($w_\sigma=0$). For training the FPs, the weights used were $w_{E} = 1.0, w_{F} = 1.0, w_{\sigma} = 0.1, w_{q} = 1.0$. For fine-tuning on the Li/Li$_6$PS$_5$Cl dataset without reference charges, $w_{q}$ was set to 0 to minimize the error of PES properties with respect to DFT calculations.

The number of training epochs was set at 1,000 and 250 for model training and fine-tuning, respectively. Early stopping was triggered when the validation loss did not decrease for 500 epochs for model training and for 50 epochs for  fine-tuning. A batch size of 32 was used to train the customized potentials for non-periodic systems and FPs. For the training of Au-MgO (001), Cu-BTA and the fine-tuning of Li/Li$_6$PSCl$_5$, a smaller batch size of 8 was used to accommodate the substantially larger system sizes in those datasets. A batch size of 4 was used for the very small molecular dimer benchmark. 

To reduce the range of target energies and improve convergence, we subtracted the total energies by the weighted sum of their elemental reference energies to arrive at a cohesive energy. These elemental reference energies, $\mathbf{E_{\mathrm{elem}}}$, were determined through linear regression according to
\begin{equation}
\mathbf{E_{\mathrm{elem}}} = (\mathbf{A}^{T}\mathbf{A})^{-1}\mathbf{A}^{T}\mathbf{E_{\mathrm{total}}},
\end{equation}
where $\mathbf{E_{\mathrm{total}}}$ denotes the vector of total energies, $N_{\mathrm{struct}}$ is the total number of structures in the dataset, $N_{\mathrm{elem}}$ is the number of distinct chemical elements, and $\mathbf{A}$ is the composition matrix constructed by stacking $\mathbf{n_{\mathrm{elem}}}$, the counts of each element per structure, across all structures. The total energy from the atomic energies can be recovered using the following expression:
\begin{equation}
    E_{total} = \sum_{i}^{N_{\mathrm{atom}}} \sigma E_{i}+\mathbf{n_{elem}} \cdot \mathbf{E_{\mathrm{elem}}}
\end{equation}
where the scaling factor $\sigma$ is the root mean square of all atomic force components in the training set.

\subsection{MD Simulations}
All molecular dynamics (MD) simulations were performed using the Atomic Simulation Environment (ASE)~\cite{larsen2017atomic} package (version 3.25.0). The time step was set to 1 fs. The Nosé-Hoover thermostat, combined with the isotropic Martyna-Tobias-Klein (MTK) barostat with a temperature damping parameter of $T_{\text{damp}}=100$ fs and a pressure damping parameter of $P_{\text{damp}}=1000$ fs, was used in the \textit{NPT} MD simulations. The Langevin thermostat with a friction coefficient of $\gamma=10$ ps$^{-1}$ was applied to perform \textit{NVT} simulations to study crystallization of phase change materials, while a much smaller coefficient of $\gamma=0.098$ ps$^{-1}$ was applied to accurately capture the dynamics of atoms in the solid electrolyte interphase.

\subsection{Benchmarking details}
For the geometry relaxation of Ag$_3^{+/-}$, each atom in the DFT structures was randomly displaced between -0.5 and 0.5 \si{\angstrom}, and the system was then relaxed using the fast inertial relaxation engine 2.0 (FIRE2) algorithm\cite{guenole2020assessment} to the nearest local minimum until the maximum force was below 0.01 eV \si{\angstrom}$^{-1}$. The PES analysis of the Au dimer in undoped and Al-doped MgO was conducted by performing 50 single-point calculations at evenly spaced intervals from 2.04~\si{\angstrom} to 2.48~\si{\angstrom}. 

The equilibrium and non-equilibrium FP benchmarks including fingerprint distance, elasticity, heat capacity were performed using MatCalc\cite{Liu_MatCalc_2024} and detailed in Ref. \citenum{kaplan2025foundational}. 
The thermal conductivity benchmark, computed using the Wigner transport equation with isotope scattering, was also performed in MatCalc using a maximum force threshold of 0.05 eV \si{\angstrom}$^{-1}$ and an atomic displacement of 0.03 \si{\angstrom}. The Li-ion migration benchmark, together with its calculation settings, was taken from the nebDFT2K dataset in Ref.\citenum{dembitskiy2025benchmarking}. The ionic conductivity benchmark was obtained from Ref.\citenum{kaplan2025foundational}.

The NaCl-\ce{CaCl2} ionic liquid test data was obtained from Ref.~\citenum{guo2023al4gap} and recomputed with the same DFT settings as the training set for the FPs. The total structure factor of the ionic liquid was computed based on the partial pair distribution functions \( g_{ij}(r) \) obtained from MD simulations. The partial structure factors \( S_{ij}(Q) \) were calculated using the Faber–Ziman formalism:
\begin{equation}
S_{ij}(Q) - 1 = \rho \int_{0}^{\infty} 4\pi r^{2}\,\bigl[g_{ij}(r) - 1\bigr]\,\frac{\sin(Qr)}{Qr}\,dr,
\label{eq:faber-ziman}
\end{equation}
where \( \rho \) is the number density, \( Q \) is the magnitude of the scattering vector, and \( r \) is the interatomic distance.
The total X-ray weighted structure factor \( S_X(Q) \) was then obtained as
\begin{equation}
S_X(Q) = \sum_{i} \sum_{j \ge i} c_i c_j f_i(Q) f_j(Q)\,\bigl[S_{ij}(Q) - 1\bigr],
\label{eq:sxf}
\end{equation}
where \( c_i \) and \( f_i(Q) \) denote the atomic concentration and X-ray scattering factor of species \( i \), respectively.
Finally, the normalized total structure factor was expressed as
\begin{equation}
S(Q) = 1 + \frac{S_X(Q)}{\left[\sum_i c_i f_i(Q)\right]^2}.
\label{eq:normalized_sf}
\end{equation}
The atomic form factors $f(s)$ were evaluated using the standard analytical expression
\begin{equation}
f(s) = Z - 41.78214\,s^{2} \sum_{n=1}^{4} a_{n}\,\exp(-b_{n}s^{2}),
\label{eq:formfactor}
\end{equation}
where $Z$ is the atomic number, $s = \sin(\theta)/\lambda$, and $a_n$ and $b_n$ are element-specific coefficients taken from the \textit{International Tables for Crystallography}.

To quantify structural evolution and crystallization kinetics during MD simulations, the SOAP descriptor\cite{bartok2012representing} implemented in DScribe\cite{himanen2020dscribe} was employed to describe the local environment of each atom. An ensemble of reference atomic environments with high crystallinity was obtained from the final snapshot from crystallization trajectories in Ref.\citenum{zhou2023device}.
The SOAP hyperparameters were set to a cutoff radius of 5.5~\si{\angstrom}, with $n_{\mathrm{max}}=8$ radial basis functions and $l_{\mathrm{max}}=4$ spherical harmonics, together with a Gaussian broadening width of $\sigma=0.5$~\si{\angstrom}.
The SOAP crystallinity metric $\bar{k}$ for each atom was computed as the average cosine similarity between its SOAP vector and the corresponding reference vectors. To increase the contrast between ordered and disordered environments, the similarity values were transformed by raising them to the fifth power ($k_{\mathrm{scaled}} = k^{5}$). The crystalline fraction over time was measured by classifying atoms with $\bar{k} > 0.7$ as crystalline. The crystalline ratio and potential energy during the time evolution were smoothed using a rolling average with a window size of 100 data points.

The radial distribution functions of the SEI for the Li/\ce{Li6PS5Cl} system were calculated from the last 100 ps of the MD trajectory using the \texttt{pymatgen-analysis-diffusion} package\cite{deng2017data}. A cutoff of 3 $\si{\angstrom}$ was set to determine the nearest neighbors of all Li atoms. For the composition profile along the $z$-axis of the simulation cell, we divided the effective simulation cell (excluding the vacuum region) into 40 bins of equal width. The fraction of lithium $f_{i, \mathrm{Li}}$ in each bin is then calculated as follows:
\begin{equation}
    f_{i,\mathrm{Li}} = \frac{N_{\mathrm{Li}}}{N_{i, \mathrm{total}}},
\end{equation}
where $N_{i,\mathrm{total}}$ denotes the total number of atoms in the volume of bin $i$. A Gaussian smoothening with $\sigma = 1.5$~\AA\ was applied. The phase classification criteria were defined as follows:
\begin{table}[htp]
    \centering
    \begin{tabular}{c|c}
        Phase &  Criteria\\
        \hline
         bulk bcc Li & $ f_{\mathrm{Li}} > 0.9$\\
         interface & $ 0.5 \leq f_{\mathrm{Li}} \leq 0.9$\\
         \ce{Li6PS5Cl} & $ f_{\mathrm{Li}} < 0.5$\\
    \end{tabular}
    \caption{Phase classification criteria for Li/\ce{Li6PS5Cl} interface simulations.}
    \label{tab:placeholder}
\end{table}

The SEI thickness was calculated for each frame of the trajectory at 10 fs intervals, and a rolling average over 200 frames (2 ps) was applied to clearly visualize trends while minimizing fluctuations.

\begin{acknowledgement}
This work was intellectually led by the U.S. Department of Energy, Office of Science, Office of Basic Energy Sciences, Materials Sciences and Engineering Division under contract No. DE-AC02-05-CH11231 (Materials Project program KC23MP). The Li/\ce{Li6PS5Cl} interface benchmarking work was supported by the Energy Storage Research Alliance (DE-AC02-06CH11357), an Energy Innovation Hub funded by the US Department of Energy, Office of Science, Basic Energy Sciences. This research used resources from the National Energy Research Scientific Computing Center (NERSC), a Department of Energy Office of Science User Facility under NERSC award DOE-ERCAP0026371. T. W. Ko also acknowledges the support of the Eric and Wendy Schmidt AI in Science Postdoctoral Fellowship, a Schmidt Futures program.

\end{acknowledgement}

\section{Data Availability}
The 4G-HDNNP datasets are available at \url{https://archive.materialscloud.org/record/2020.137}. The Li/Li$_6$PS$_5$Cl dataset is available at \url{https://aissquare.com/datasets/detail?pageType=datasets&name=Training_data&id=347}. The weights of all FPs trained in this work are available directly via Hugging Face at \url{https://huggingface.co/materialyze/models}.

\section{Code Availability}
The QET architecture is implemented in the open-source Materials Graph Library (MatGL) package (\url{https://github.com/materialyzeai/matgl}). The DIRECT sampling algorithm is implemented in the Materials Machine Learning (maml) package (\url{https://github.com/materialyzeai/maml}), 
and the FP benchmarks were performed using the MatCalc package (\url{https://github.com/materialyzeai/matcalc}).

\section{Author contributions}
T.W.K. and S.P.O. conceived the idea and designed the study. T.W.K. proposed and implemented the QET architectures and supervised the project from data generation to the design of benchmarks. R.L. led the benchmarking for FPs, while A.R.M. led the benchmarking for fine-tuning MLIPs and conducted dimensionality reduction analyses. R.L., Z.Y., and J.Q. contributed to data sampling and generation. T.W.K. and S.P.O. prepared the initial draft, and all authors contributed to the discussion and revision.

\section{Completing Interests}
A provisional patent application has been filed covering aspects of the QET architecture and training. 

\bibliography{Bibliography}
\includepdf[pages=-]{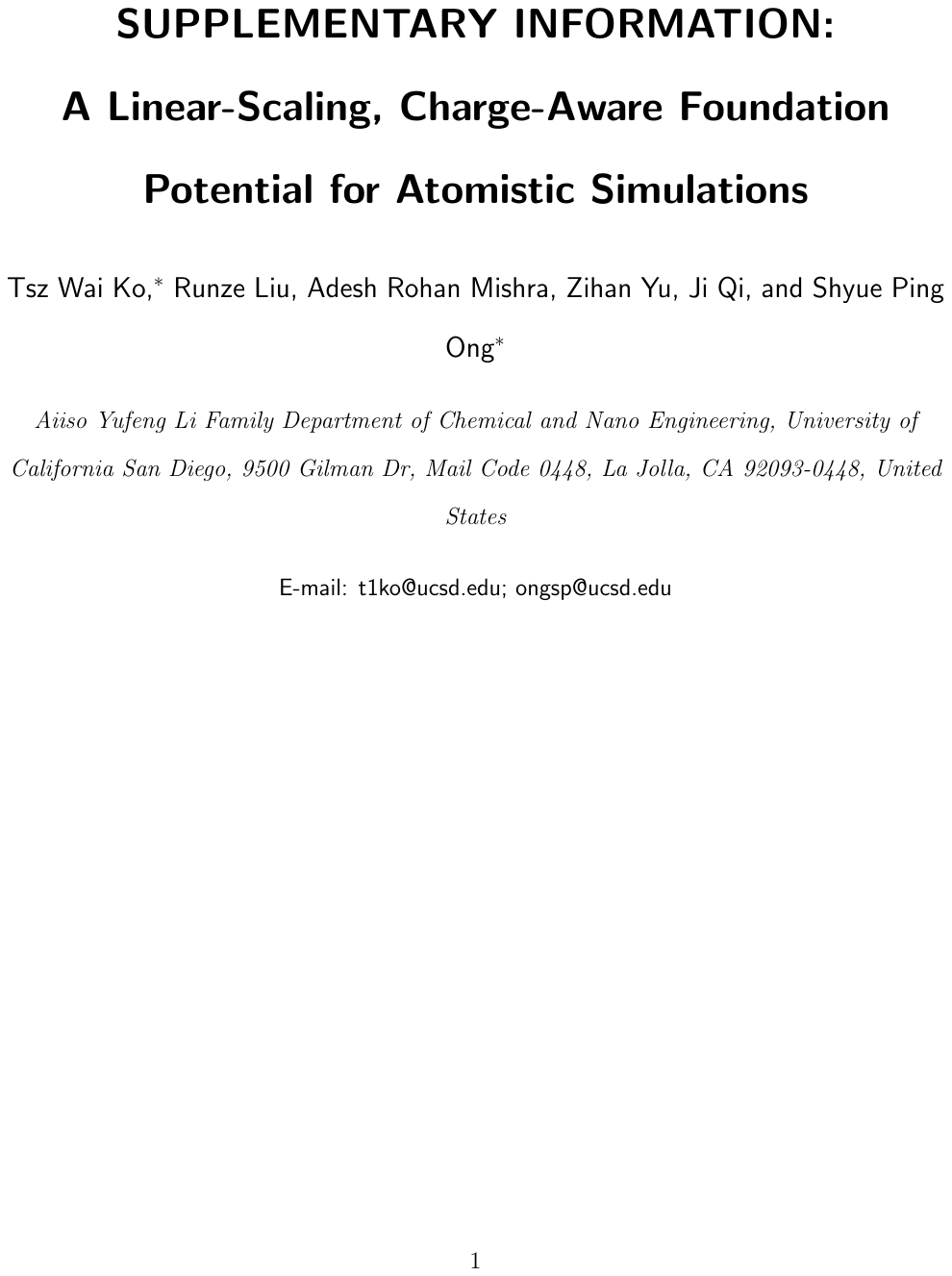}
\end{document}